\documentclass[twocolumn, amssymb, amsmath, aps, prd, showpacs, nofootinbib]{revtex4}

\usepackage{amssymb}
\usepackage{amsmath}
\usepackage[usenames]{color}
\ifx\pdfoutput\undefined
\usepackage{graphicx}
\else
\usepackage[pdftex]{graphicx}
\usepackage{epstopdf}
\usepackage{mathrsfs}
\fi
\usepackage[usenames]{color}
\definecolor{Black}{rgb}{1.,0.,0.}

\def\reff@jnl#1{{\rm#1\/}}
\def\aj{\reff@jnl{AJ}}                	 
\def\jcap{\reff@jnl{JCAP}} 

\usepackage{hyperref}

\providecommand{\adsurl}[1]{\href{#1}{ADS}}

\providecommand{\url}[1]{\href{#1}{#1}}

\newcommand{\scsc}[1]{{\scriptscriptstyle{#1}}} 

\newcommand{\OmegaL}{\ensuremath{\Omega_{\scsc{\Lambda}}}}

\newcommand{\dif}{\mathrm{d}}
\newcommand{\himpc}{\ensuremath{\,h\,{\rm Mpc}^{-1}}}

\def\sub#1{_{\mbox{\scriptsize{#1}}}}

\providecommand{\adsurl}[1]{\href{#1}{ADS}}

\providecommand{\url}[1]{\href{#1}{#1}}

\newcommand{\fref}[1]{FIG.~\ref{#1}}
\newcommand{\tref}[1]{TABLE~\ref{#1}}
\newcommand{\cref}[1]{Chapter~\ref{#1}}
\newcommand{\sref}[1]{Section~\ref{#1}}

\newcommand\eq[1]{Eq.~(\ref{#1})}

\newcommand{\beq}{\begin{equation}}
\newcommand{\eeq}{\end{equation}}
\newcommand{\bea}{\begin{eqnarray}}
\newcommand{\eea}{\end{eqnarray}}

\newcommand{\lsim}{\,\raise 0.4ex\hbox{$<$}\kern -0.8em\lower 0.62ex\hbox{$\sim$}\,}

\begin{document}

\title{Large Scale Structure Forecast Constraints on Particle Production During Inflation}
\author{Teeraparb Chantavat}
\email{txc@astro.ox.ac.uk}
\affiliation{Oxford Astrophysics, Denys Wilkinson Building, Keble Road, Oxford, OX1 3RH, United Kingdom.}
\author{Christopher Gordon}
\affiliation{Oxford Astrophysics, Denys Wilkinson Building, Keble Road, Oxford, OX1 3RH, United Kingdom.}
\author{Joseph Silk}
\affiliation{Oxford Astrophysics, Denys Wilkinson Building, Keble Road, Oxford, OX1 3RH, United Kingdom.}

\begin{abstract}

Bursts of particle production during inflation provide a well-motivated mechanism for creating bump-like features in the primordial power spectrum. Current data constrains these features to be less than about 5\% the size of the featureless primordial power spectrum at wavenumbers of about $0.1 \himpc$. We forecast that the Planck cosmic microwave background experiment will be able to strengthen this constraint to the 0.5\% level. We also predict that adding data from a square kilometer array (SKA) galaxy redshift survey would improve the constraint to about the 0.1\% level. For features at larger wave-numbers, Planck will be limited by Silk damping and foregrounds. While, SKA will be limited by non-linear effects. We forecast for a Cosmic Inflation Probe (CIP)  galaxy redshift survey, similar constraints can be achieved up to about a wavenumber of $1.0 \himpc$.

\end{abstract}

\maketitle

\section{Introduction}

Current data is remarkably consistent with predictions of the simplest models of inflation. To a high degree of accuracy, the Universe appears to be flat and have nearly scale-invariant, isotropically distributed, Gaussian, and adiabatic primordial perturbations \cite{Bennett_ea2010, Komatsu_ea2010}. However, there are still potentially large improvements to be implemented  in the precision and the length scales of the primordial perturbations that will be probed. Therefore, it is important to investigate possible deviations from this simple picture that may in future be detectable. In this article, we concentrate on features in the primordial power spectrum that may be caused by bursts of particle production during inflation \cite{Barnaby_Huang2009, Barnaby_ea2009, Barnaby2010}. This can happen when the motion of the inflaton causes the mass of another field to pass through zero. The resultant production of particles leads to a corresponding bump like  feature in the primordial power spectrum at around the scales which are then  leaving the Hubble horizon.

A feature in the primordial power will translate to a corresponding feature in the cosmic microwave background  (CMB) angular power spectrum and the matter power spectrum. The matter power spectrum may be probed  in many ways, and in this article we concentrate on galaxy redshift surveys and cluster surveys.

Currently, there is no detection of such features in the data, but only wave-numbers of $k \lesssim 0.1\himpc$ have been probed and only to accuracies of about 5\% \cite{Barnaby_Huang2009}. The Planck CMB survey can probe smaller length scales 
due to higher resolution and lower noise and so will help improve the constraints
up to to $k\sim 0.2 \himpc$ where noise, beam size limitations and foregrounds start to dominate. In general astro-physical foregrounds prevent one probing the primordial power spectrum beyond about $k\sim 0.5 \himpc$ using the CMB.
Future galaxy redshift surveys such as are planned with the Square Kilometer Array\footnote{See http://www.skatelescope.org/} (SKA), will also probe $k\sim 0.1\himpc$ to much higher accuracy due to the huge volumes that they will encompass
but will be limited to $k\lesssim 0.2\himpc$ by non-linear effects.
While,  $k\lesssim2\himpc$ may be probed by very high redshift galaxy surveys, such as the Cosmic Inflation Probe\footnote{See http://www.cfa.harvard.edu/cip/} (CIP), where non-linear growth has yet to dominate.

In this article, we will make forecasts on how well the amplitude of a particle production-induced feature can be constrained by future large scale structure surveys. We begin in \sref{sec:feature} with a summary of how particle production  can generate a bump-like feature. In \sref{sec:fisher}, we give an overview of Fisher information matrices, a conventional method of predicting constraints on a set of parameters, for CMB, galaxy and cluster surveys. Predictions for constraints on the  feature amplitude and position combined with other cosmological parameters are given in \sref{sec:forecast}. We discuss our results in \sref{sec:discussion}.

\section{Particle Production}
\label{sec:feature}

\begin{figure}\centering
	\includegraphics[scale=0.45, viewport=400 0 200 400]{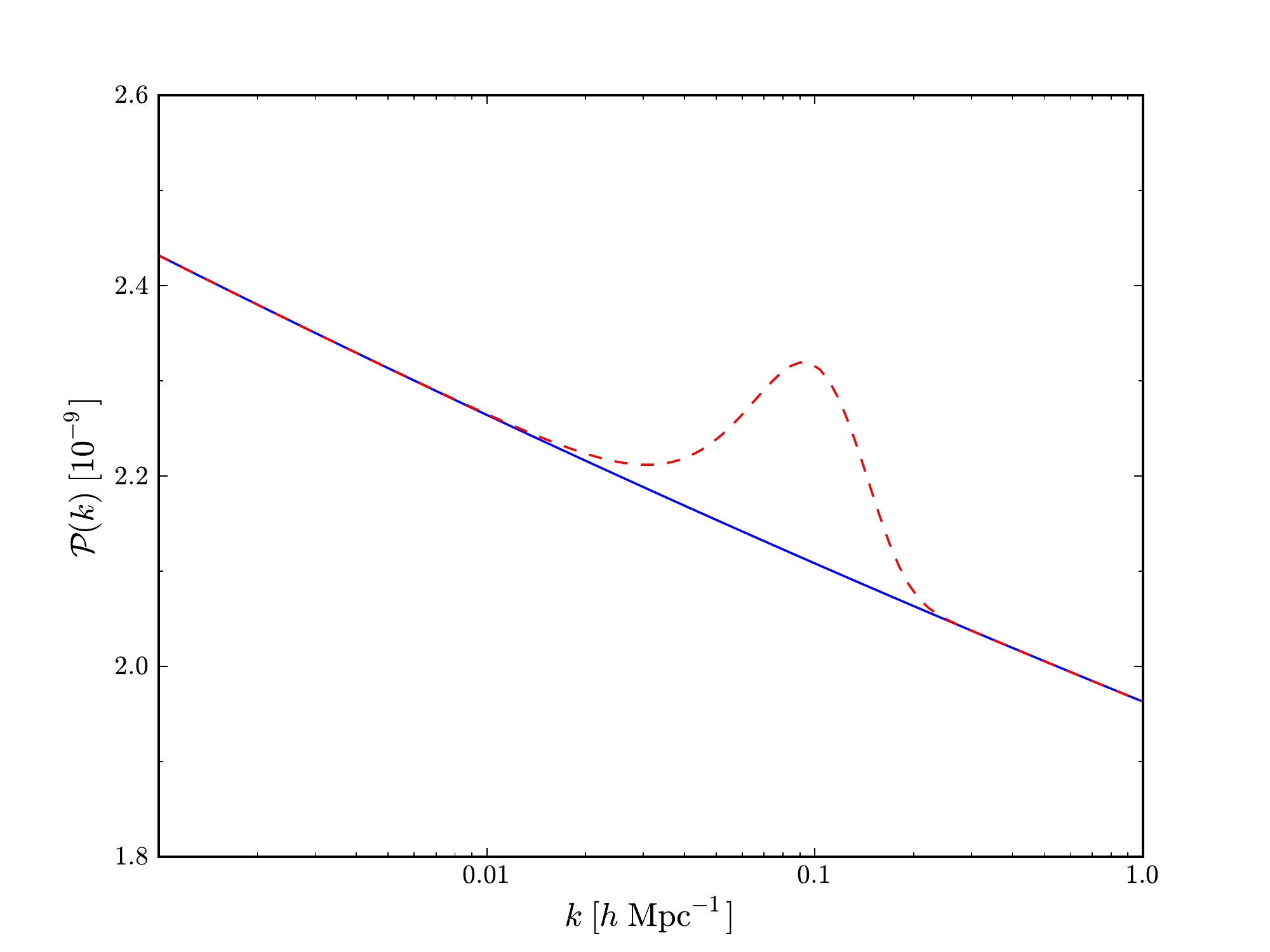}
	\caption{\label{fig:feature} The primordial power spectrum plus  particle \newline creation features with $A\sub{\tiny IR}$ = 1.25 $\times 10^{-10}$ at position \newline $k\sub{\tiny IR}= 0.1 \himpc$ (red dashed).}
\end{figure}

Recently, a mechanism has been proposed that will generate a bump-like feature through particle production during inflation \cite{Barnaby_ea2009, Barnaby_Huang2009, Barnaby2010}. In this scenario, the production of massive iso-inflaton particles during inflation gives rise to potentially quantitatively observable features in the primordial power spectrum. The fields simply interact via the coupling
\begin{equation}
	\mathcal{L}\sub{int} = -\frac{g\sub{\tiny IR}^2}{2}(\phi - \phi_\scsc{0})^2\chi^2,
\end{equation}
where $g\sub{\tiny IR}$ is the interaction coupling constant. $\phi$ and $\chi$ are the inflaton and iso-inflaton fields respectively. When $\phi$ passes through $\phi_\scsc{0}$ there is a non-adiabatic change in the mass of $\chi$ and as a result a burst of particle production. This rapidly drains energy from the $\phi$ field which can lead to a transient violation of slow roll and hence an associated  ``ringing'' in the primordial curvature fluctuations which is similar to that seen in models with a sharp feature in the potential, \cite{Adams_ea2001, Chen_ea2007, Chen_ea2008, Hunt_Sarkar2004, Hunt_Sarkar2007, Joy_ea2008, Lerner_McDonald2009, Mortonson_ea2009, Starobinskij1992}. However, the dominant effect is found to come from  multiple re-scatterings of the produced $\delta \chi$ particles off the $\phi$ condensate. Multiple bump-like features scenario is also possible with $\phi_\scsc{0}$'s in different positions, which associates with different $g\sub{\tiny IR}$'s. However, we restrict to the case of only one feature presents. The overall effect is a bump-like feature which can approximately be described by a parametric form as
\begin{gather}
	\label{eq:psIRfeature}
	\mathcal{P}(k) = \hspace{10cm} \\
	\Delta_\scsc{\mathcal{R}}^2 \left( \frac{k}{k\sub{pivot}} \right)^{n_\scsc{s} - 1} + A\sub{\tiny IR} \left( \frac{\pi e}{3} \right)^{3/2}\left(\frac{k}{k\sub{\tiny IR}}\right)^3 e^{-\frac{\pi}{2}\left( \frac{k}{k\sub{\tiny IR}}\right)^2}, \nonumber
\end{gather}
where $\Delta_\scsc{\mathcal{R}}^2$ is the scalar amplitude describing the normalisation of the power spectrum. $n_\scsc{s}$ and $k\sub{pivot}$ is the tilt and the pivot wavenumber respectively. The first term on the RHS of \eq{eq:psIRfeature} is the standard power-law power spectrum and the second term contains the features generated by particle creation which is parameterised by an amplitude, $A\sub{\tiny IR}$ and position $k\sub{\tiny IR}$ as shown in \fref{fig:feature}. The normalisation is chosen so that $A\sub{\tiny IR}$ is the amplitude of the feature at its peak $k\sub{\tiny IR}$. The relation to the coupling constant is given by
\begin{equation}
	A\sub{\tiny IR} \approx 1.01\times10^{{-6}}g\sub{\tiny IR}^{15/4}\,.
\end{equation}

\section{Fisher Matrix Calculation}
\label{sec:fisher}

We use the Fisher Information Matrix \cite{Tegmark_ea1997} to make predictions for constraints of cosmological parameters for future surveys. The statistics that will be implemented are cluster number counts, the cluster/galaxy power spectrum and the cosmic microwave background power spectrum.

\subsection{Cluster Number Count}

The simplest statistics we can extract from a cluster survey is the number count. 

\subsubsection{Differential Halo Mass Function}

In order to predict the number density of collapsed objects in the Universe, a statistical concept of halo or mass distribution is used here. The \textit{differential halo mass function}, or \textit{halo mass function} for short, is defined as the redshift-dependent distribution of the number of collapsed dark matter haloes per unit mass interval in a unit co-moving volume. The halo mass function is given by
\begin{equation}
	\label{eq:massfunction}
	\frac{\mbox{d}n}{\dif M} = \frac{\rho_\scsc{m}}{M} \frac{\dif\ln\sigma^{-1}}{\dif M} f(\sigma),
\end{equation}
where $\mbox{d}n/\dif M$ is the differential halo mass function, $\rho_\scsc{m}$ is the matter density, and  $f(\sigma)$ is called the \textit{mass fraction}. The smoothed variance is calculated as 
\begin{equation}
	\label{eq:sigma}
	\sigma^2(R ,z) = \frac{D(z)^2}{2\pi^2} \int_0^{\infty} P(k)\ W^2(k, R)\ k^2\dif k,
\end{equation}
where $D(z)$ is the linear growth function normalised to 1 at the present epoch. $W(k, R)$ is the Fourier-space top-hat window function. $R$ is the  smoothing radius for a comoving sphere enclosing a mass of 
\begin{align}
	\label{eq:massscale}
	M &={4\pi\over 3} R^3 \rho_\scsc{m} \nonumber \\ &=1.16\times10^{12}\Omega_\scsc{m} \left({R \over h^{-1}{\rm Mpc}} \right)^{3} h^{-1}M_\scsc{\odot}\, .
\end{align}
The top hat smoothing in \eq{eq:sigma} suppresses the contribution of any change to the primordial power spectrum located at wave number $k\sub{\tiny IR} \gg 1/R$. Combined with \eq{eq:massscale}, this implies that a change in the primordial power spectrum at $k\sub{\tiny IR}$ has a suppressed effect on the number density on mass scales satisfying
\begin{equation}
	{M \over h^{-1}M_\scsc{\odot}} \gg 10^{12}  \left ({k\sub{\tiny IR} \over \himpc}\right)^{-3} \, .
\end{equation}

The \textit{mass fraction}  is defined as a fraction of mass in collapsed haloes per unit interval in $\ln\sigma^{-1}$. The halo mass function is described by a pair of parameters, $f(\sigma)$ and $\ln\sigma^{-1}$. Both of these parameters are a natural way of parameterising the mass function from different cosmological models with the fewest number of parameters. All the cosmological parameters are embedded in $\sigma$. $f(\sigma)$ is the fraction of the density fluctuation that eventually collapse into non-linear objects. We used the mass function by Jenkins \textit{et al.} (2001) \cite{Jenkins_ea2001}:
\begin{equation}
	f(\sigma) = 0.315 \exp\left[ -|\ln\sigma^{-1} + 0.61|^{3.8} \right].
\end{equation}

\subsubsection{Number Count Fisher Matrix}

For a survey which covers $f\sub{sky}$ fraction of the sky, a theoretically expected value of the number of clusters in the  $i$th  redshift bin at central redshift $z_i$ and a width of $\Delta z$ is given by
\begin{gather}
	\label{eq:NCount}
	\bar{N}(M > M\sub{lim}, z) = \hspace{5cm} \\ f\sub{sky} \int_{z_i - \frac{1}{2} \Delta z}^{z_i + \frac{1}{2} \Delta z} \dif z\frac{\dif V}{\dif z} \int_{M\sub{lim}}^\infty\dif M \frac{\dif n}{\dif M}(M, z). \nonumber
\end{gather}
The differential co-moving volume, $\dif V/\dif z$, is
\begin{equation}
	\frac{\dif V}{\dif z} = \frac{4\pi}{H(z)} \left[ \int_0^z \frac{\dif z^\scsc{\prime}}{H(z^\scsc{\prime})} \right]^2,
\end{equation}
where $H(z)$ is the Hubble parameter at redshift $z$. \eq{eq:NCount} includes all the clusters above a threshold mass of $M\sub{lim}$. The mass threshold can, in general, be redshift-dependent which is normally given in terms of a survey-specific \textit{selection function}. To investigate generally how cluster surveys can be used to probe features, we will use a simple redshift-independent effective mass threshold for cluster surveys which yields an equivalent number of clusters over the entire survey volume. 

Given a set of parameters of interest, $\Theta = (\theta_\scsc{1}, \theta_\scsc{2}, ..., \theta_\scsc{m})$, an element $\theta_\scsc{\mu}$ and $\theta_\scsc{\nu}$ of a Fisher matrix for number count is \cite{Holder_ea2001}
\begin{equation}
	F_\scsc{\mu\nu} = \sum_{i = 1}^{N\sub{bins}} \frac{1}{\bar{N}_i}\frac{\partial \bar{N}_i}{\partial\theta_\scsc{\mu}}\frac{\partial \bar{N}_i}{\partial\theta_\scsc{\nu}},
\end{equation}
where $F_\scsc{\mu\nu}$ is a sum of all redshift bins in the survey.

\subsection{Power Spectrum}

We also consider cosmological constraints from measurements of the matter power spectrum.

\subsubsection{Galaxy Power Spectrum}

Galaxies are moving away from us along with the Hubble flow. The perturbation of the velocity of the galaxies is called the \textit{peculiar velocity} which is the motion of the galaxies relative to the Hubble flow. It changes the observed Doppler shift and, hence, the distance inferred from redshift measurement.  Kaiser (1987) \cite{Kaiser1987} showed that the power spectrum derived from a redshift survey, $P_\scsc{s}(\mathbf{k})$, is given by
\begin{equation}
	P_\scsc{s}(k, \mu) = \left[ 1 + \beta\mu^2 \right]^2 b^2 P(k),
\end{equation}
where $P(k)$ and $P_\scsc{s}(k)$ are the matter power spectrum and the redshift power spectrum respectively, $\mu \equiv \widehat{\mathbf{k}} \cdot \widehat{\mathbf{n}}$ is the cosine of the angle between $\mathbf{k}$ and the line of sight,  $\beta$ is
\begin{equation}
	\beta = \frac{1}{b} \frac{\dif\ln D}{\dif\ln a},
\end{equation}
where $D$ is the linear growth function normalised to 1 at the present epoch, and $b \equiv \delta_\scsc{g} / \delta$ is the galaxy bias. The bias for galaxy surveys is normally estimated from the data and the uncertainty in bias is propagated into the other parameter constraints. {  However, to include non-linear effects, we applied a Taylor-like expansion to the bias as
\begin{equation}
	b^2 = b_0^2 \left[ 1 + a_\scsc{1} k + a_\scsc{2}  k^2 \right],
\end{equation}
where $b_0$ is a scale independent bias. $a_\scsc{1}$ and $a_\scsc{2}$ are the first and the second order term respectively \cite{ReidPercivalEisenstein:2010}.
}

\begin{figure}\centering
	\includegraphics[scale=0.45]{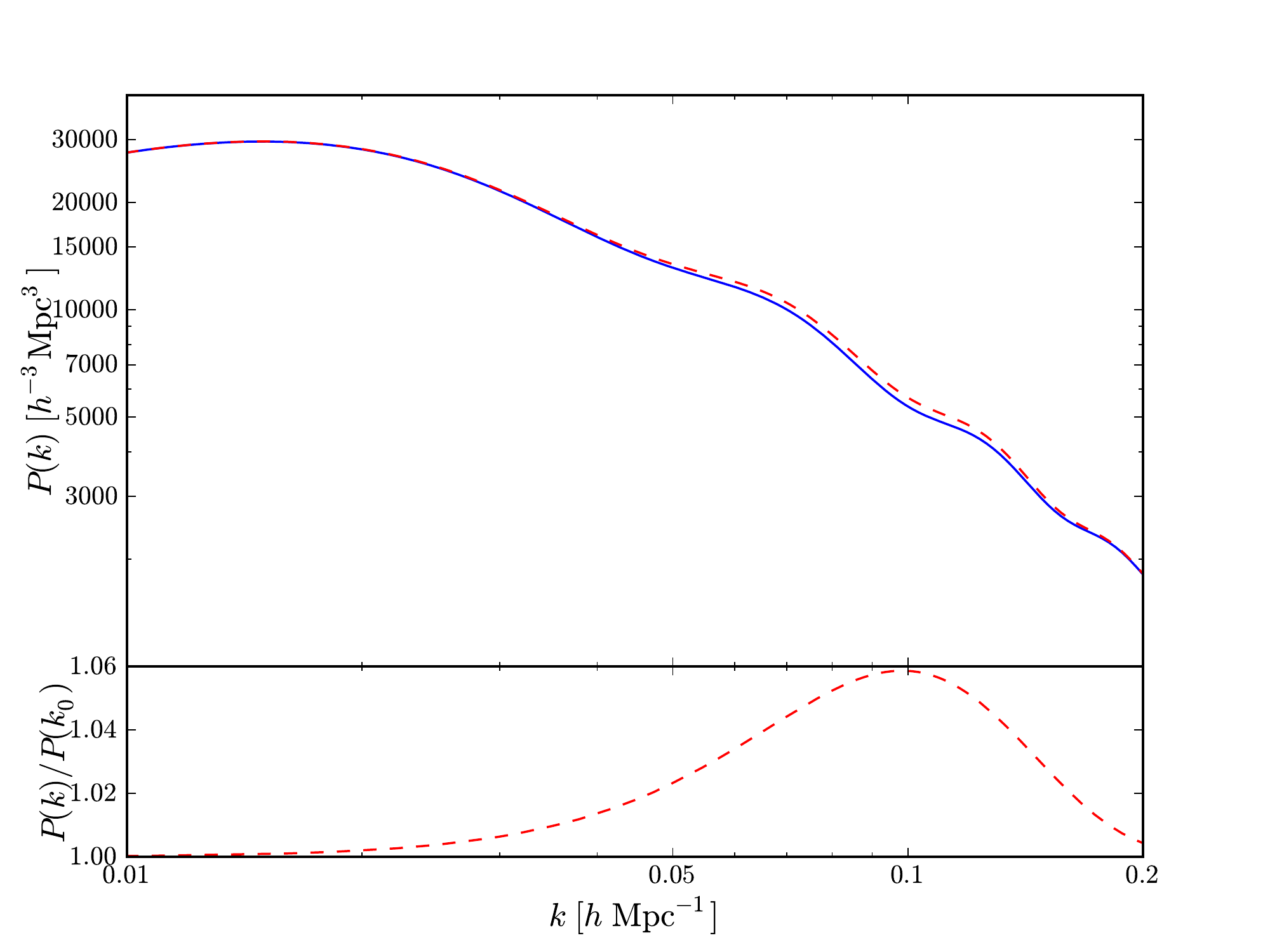}
	\caption{\label{fig:matterf} The matter power spectrum from a featureless primordial power spectrum (blue solid) and the matter power spectrum from a primordial power spectrum  which has  a particle creation feature  with amplitude  $A\sub{\tiny IR}$ = 1.25 $\times 10^{-10}$ at position $k\sub{\tiny IR} = 0.1 \himpc$ (red dashed).}
\end{figure}

The linear theory matter power spectrum is related to the primordial power spectrum by
\begin{equation}
	P(k)\propto T^2(k)\ k\ {\cal P}(k)
\end{equation}
where $T$ is the transfer function. So a narrow feature in the primordial power spectrum will be transferred to a narrow feature in the matter power spectrum at about the same comoving wavenumber, see \fref{fig:matterf}. 

\subsubsection{Cluster Power Spectrum}

The Poisson approximation for cluster number counts is not strictly accurate for larger surveys \cite{HuKravtsov:2003}.
There is an additional sample variance which can be used to help constrain the mass scaling relation. We account for 
this by incorporating an additional constraint from the power spectrum of the clusters.
The way of doing this is similar to that of galaxies. However, in the case of cluster bias, we use an effective linear halo bias
\begin{equation}
	\label{eq:biaseff}
	b\sub{eff}(z) = \frac{\int_{M\sub{lim}}^\infty\dif M\ b(M, z) \dif n/\dif M}{\int_{M\sub{lim}}^\infty\dif M\ \dif n/ \dif M},
\end{equation}
where $\dif n/\dif M$ is the differential mass function (See \eq{eq:massfunction}) and $b(M, z)$ can be calculate from halo bias \cite{Sheth_Tormen1999}
\begin{equation}
	b(\nu) = 1 + \frac{a\nu - 1}{\delta_\scsc{c}} + \frac{2p}{\delta_\scsc{c} [1 + (a\nu)^p]},
\end{equation}
where $\nu = (\delta_\scsc{c} / \sigma)^2$ with $\delta_\scsc{c} = 1.69$, $a = 0.75$ and $p = 0.3$.

\subsubsection{ Power Spectrum Fisher Matrix }

We may write the appropriate Fisher Matrix for the power spectrum by assuming the likelihood function to be Gaussian as \cite{Tegmark1997}
\begin{equation}
	\label{eq:PSfisher}
	F_\scsc{\mu\nu} = \frac{1}{2} \int_{k\sub{min}}^{k\sub{max}} \frac{\dif^3 \mathbf{k}}{(2\pi)^3} \frac{\partial\ln P_\scsc{s}(\mathbf{k})}{\partial\theta_\scsc{\mu}} V\sub{eff}(\mathbf{k}) \frac{\partial\ln P_\scsc{s}(\mathbf{k})}{\partial\theta_\scsc{\nu}}\, .
\end{equation}
We set $k\sub{max}$ to be the wavenumber where non-linear effects start to become non-negligible. From \cite{Seo_Eisenstein2003} where $\sigma(R\sub{nl}) = 0.5$ and $k\sub{max}=k\sub{nl} = \pi/2/R\sub{nl}$. We set $k\sub{min} = 1.0 \times 10^{-4}$ Mpc$^{-1}$ for the lower limit. The effective survey volume, $V\sub{eff}$, is  a given by
\begin{eqnarray}
	\label{eq:veff}
	\nonumber V\sub{eff}(k, \mu) & = & \int \dif^3\mathbf{r}\left[\frac{n(\mathbf{r}) P_\scsc{s}(k, \mu)}{n(\mathbf{r}) P_\scsc{s}(k, \mu) + 1} \right]^2 \\
	& \approx & \left[ \frac{\bar{n} P_\scsc{s}(k, \mu)}{\bar{n} P_\scsc{s}(k, \mu) + 1} \right]^2 V\sub{survey}.
\end{eqnarray}
where $n(\mathbf{r})$ is the number density at position $\mathbf{r}$ and $\bar{n}$ is the average number density for the survey. The effective volume is due to a finite survey volume and incomplete sampling of the underlying density field. These are known as sample variance and shot noise respectively. The uncertainty is propagated through the calculation by the weighing factor $[n(\mathbf{r}) P_\scsc{s}/( n(\mathbf{r}) P_\scsc{s} + 1)]^2$. For galaxy surveys where the number density is high ie. $\bar{n} P_\scsc{s}(k, \mu) \gg 1$, the effective survey volume is then $V\sub{eff} \approx V\sub{survey}$.

\subsection{CMB}


The  primordial power spectrum is probed over a wide range of wave numbers by measurements of the primary CMB anisotropies  (see for example \cite{Hu_Okamoto2004}). Both the temperature ($T$) and $E$-mode of the polarization ($E$) probe scalar perturbations. 
\begin{equation}
	{\ell(\ell+1)C_\scsc{\ell}^\scsc{X} \over 2\pi} = \int \mbox{d} \ln k\, (T^\scsc{X}_\scsc{\ell}(k))^2 {\cal P}(k),
\end{equation}
where $X \in T, E$ for auto-correlation function ($TT$, $EE$). For the cross-correlation power spectrum ($TE$),
\begin{equation}
	{\ell(\ell+1)C_\scsc{\ell}^\scsc{C} \over 2\pi} = \int \mbox{d} \ln k\, T^\scsc{T}_\ell(k)T^\scsc{E}_\ell(k) {\cal P}(k).
\end{equation}

\begin{figure}\centering
	\includegraphics[scale=0.45]{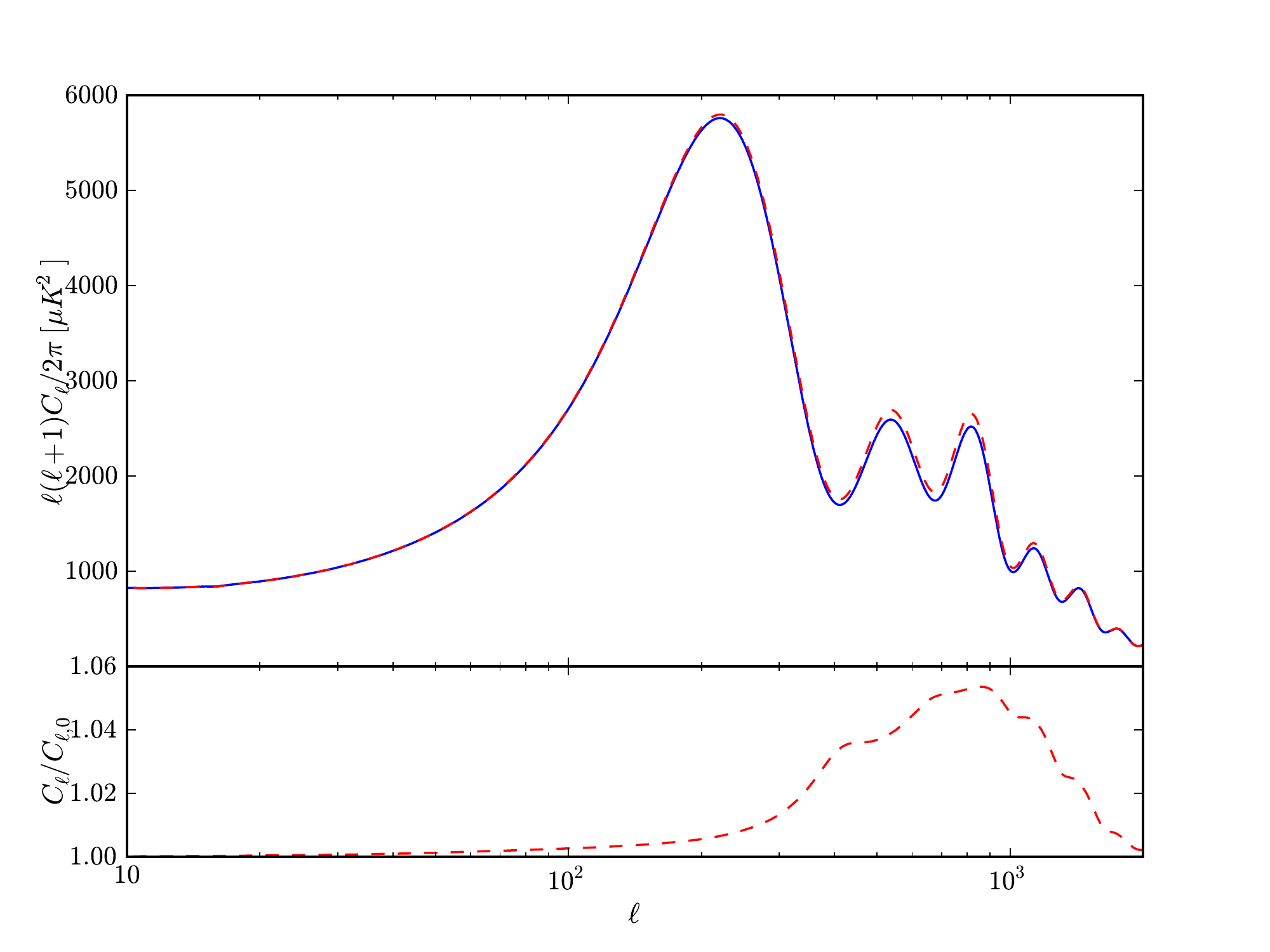}
	\caption{\label{fig:cmbf} The CMB angular power spectrum from a featureless primordial power spectrum (blue solid) and the CMB angular power spectrum  from a  primordial power spectrum with a particle creation feature which has $A\sub{\tiny IR}$ = 1.25 $\times 10^{-10}$ at position $k\sub{\tiny IR}= 0.1 \himpc$  (red dashed).}
\end{figure}

The projection of a mode of wave-number $k$ on to the surface of last scattering  (a sphere of comoving radius $D_*$) results in the CMB transfer functions having the form $T^X_\ell \sim j_\ell(kD_*)$. Where $j_\ell$ is the spherical Bessel function of order $\ell$ which peaks at $\ell\approx kD_*$. Therefore a bump in the primordial power spectrum at wave-number $k_i$ is mapped onto a bump in CMB angular power spectrum at
\begin{equation}
	\ell\sim k\sub{\tiny IR} D_*\approx 10^4 {k\sub{\tiny IR} \over h {\rm Mpc}^{-1}},
\end{equation} 
see \fref{fig:cmbf}

The foreground contribution from secondary sources which will probably be hard to completely remove for $\ell>2000$ for both temperature and polarization. For this reason, as done by \cite{Hu_Okamoto2004} and \cite{Leach2006}, we will restrict ourselves to $\ell\leq2000$ when evaluating the forecasted marginalized errors.

The CMB  Fisher matrix is given by (see for example \cite{Zaldarriaga_ea1997})
\begin{equation}
	{F}_\scsc{ij}=\sum_{\ell} \sum_{X,X'} {\partial C_\scsc{\ell}^\scsc{X} \over \partial \theta_\scsc{i}} {\rm Cov}^{-1} (C_\scsc{\ell}^\scsc{X}, C_\scsc{\ell}^\scsc{X'}){\partial C_\scsc{\ell}^\scsc{X'} \over \partial \theta_\scsc{j}}
\end{equation}
where the covariance matrix can be obtained from \cite{Zaldarriaga_ea1997} and it depends on the temperature  noise per pixel ($\sigma_\scsc{T}$), the polarization noise  per pixel ($\sigma_\scsc{E}$), the pixel area in radians squared ($\theta^2 = 4\pi/N_{\rm pix}$), and the beam window function which we approximate as Gaussian 
($B_\scsc{\ell}\approx \exp(-\ell(\ell+1)\sigma_\scsc{b}^2$). 

\section{Forecast Constraints}
\label{sec:forecast}

With the descriptions of Fisher information matrices in \sref{sec:fisher}, we can make predictions for future upcoming surveys. We also test our Fisher matrix formalism on some current surveys and compare to published results. For the CMB we take the Planck\footnote{http://www.rssd.esa.int/index.php?project=planck} and WMAP\footnote{http://map.gsfc.nasa.gov/} surveys. The values for Planck are taken from the Planck blue book\footnote{http://www.rssd.esa.int/SA/Planck/docs/Bluebook-ESA-SCI(2005)1\_V2.pdf} and are listed in \tref{table:planckparam} (note that $\theta$ needs to be converted to radians).
\begin{table}\centering
	\caption{\label{table:planckparam} Planck Instrument Characteristics}
	\begin{tabular}{|l|c|c|c|c|}
		\hline
		Center Frequency (GHz) & 70 &100 &143 & 217 \\
		\hline\hline
		$\theta$ (FWHM arcmin) &14 &10 & 7.1 & 5.0 \\
		$\sigma_\scsc{T}$ ($\mu{\rm K}$)& 12.8 & 6.8 & 6.0 & 13.1\\
		$\sigma_\scsc{E}$ ($\mu{\rm K}$)& 18.2 & 10.9 & 11.4 & 26.7 \\
		\hline
	\end{tabular}
\end{table}
We use $\sigma_\scsc{b}=\theta/\sqrt{8 \log 2}$ and combine the different frequency bands as specified in \cite{Bond_ea1997}.  We take the range in $\ell$ to be 2 to 2000. At higher $\ell$, secondary sources of temperature and polarization will likely prohibit the extraction of cosmological information from the primary CMB. 

For galaxy surveys, we consider the Cosmic Inflation Probe\footnote{See http://www.cfa.harvard.edu/cip/} (CIP) \cite{Melnick_ea2004}, which is a space-based mission aimed to measure the linear galaxy power spectrum over $k \sim 0.03 - 2.0 \himpc$ to better than 1\%. The primary science goal of the CIP is to provide constraint on inflation models by observing H$\alpha$. The CIP survey will cover 1,000 square degrees and will be capable of detecting more than $10^8$ galaxies between redshift range of 1.8 - 6.5. We follow the CIP model from by having redshift bins at $z = 2.0 - 3.5,\ 3.5 - 5.0,\ 5.0 - 6.5$ and and having an average galaxy number density of 1.0 $\times 10^{-2}$, 5.3 $\times 10^{-3}$ and 1.3 $\times 10^{-3}$ $h^3$ Mpc$^{-3}$ respectively which is equivalent to $\sim$100 million of galaxies within the survey volume of $\sim$15.0 $h^{-3} \mbox{Gpc}^3$. We restrict our calculation to $k\sub{max} = 0.5,\ 1.0,\ \mbox{and}\ 2.0 \himpc$ for the three bins respectively (See \eq{eq:PSfisher}).

The Sloan Digital Sky Survey\footnote{See http://www.sdss.org/} (SDSS) \cite{York_ea2000} is a ground-based optical survey using a 2.5 meter telescope at Apache Point Observatory, New Mexico. It final data set (DR7) includes more than 230 million celestial objects and spectra of about 930,000 galaxies and 120,000 quasars over an area of 8,400 square degree in five optical bandpasses to redshift about 1.0. We follow the SDSS survey model from Pritchard \& Pierpaoli (2008) \cite{Pritchard_Pierpaoli2008}. We estimate the bias for the SDSS survey as 2.25 (ie. $b \equiv \sigma_{8, g} / \sigma_8$ where $\sigma_{8, g} = 1.8$ and $\sigma_8 = 0.8$) and $\bar{n} = 1.0\ \times\ 10^{-4}\ h^3 \mbox{Mpc}^{-3}$. The Square Kilometre Array\footnote{See http://www.skatelescope.org/} (SKA) is a large-scale radio telescope which  aims to cover the frequency range from 60 MHz to 35 GHz and 20,000 square degree of the sky. The main science goal of the SKA project is to study the HI content of galaxies to cosmologically significant distances, $z \sim 2.0$, and to make reionisation maps using 21 cm transition of neutral hydrogen. The SKA project is currently in a design phrase and the telescope site will be decided in 2012. We consider the galaxy redshift survey component of SKA (see for example \cite{Abdalla_ea2010}). It also may be possible to use the 21 cm absorption component of the SKA to probe reionization and matter power spectrum at a higher redshift (see for example \cite{Adshead_ea2010}). We shall consider the SKA survey as a stereotype of a cosmic variance-limited galaxy redshift survey by assuming that the number density is so high  that $\bar{n} P_s \gg 1.0$. Hence, the effective survey volume is equivalent to the survey volume (see \eq{eq:veff}) from $z = 0.0 - 2.0$. The set of parameter of all the galaxy surveys will include bias as an additional parameter reflecting the fact that we do not know accurately the value of the bias. The uncertainty in bias will propagate into the parameter constraints.
\begin{table*}[h]\centering
	\caption{\label{table:surveys} Details for galaxy and cluster surveys.}
	\begin{tabular}{|l|c|c|c|c|c|c|c|}
		\hline
		Survey &\ \ \ \ \ \ $z$\ \ \ \ \ \ & \ \ \ \ $\Delta z\sub{bin}$\ \ \ \ & $V\sub{survey}$ & $k\sub{max}$  & \ \ \ \ $b$\ \ \ \ \ & $M\sub{lim}$ & $f\sub{sky}$ \\
		& & & ($h^{-3}$ Gpc$^3$) & ($\himpc$) & & ($h^{-1} M_\odot$) & \\
		\hline\hline
		\underline{Galaxy Survey} & & & & & & & \\
		CIP & 2.0 -- 6.5 & 1.5 & 15.0 & 2.0 & 1.0 & N/A & 0.024 \\
		SDSS & 0.0 -- 0.6 & 0.6 & 1.0 & 0.1 & 2.25 & N/A & 0.3 \\
		SKA & 0.0 -- 2.0 & 2.0 & 100 & 0.4 & 1.0 & N/A & 0.5 \\
		\hline
		\underline{Cluster Survey} & & & & & ($b\sub{eff}$) & & \\
		SNAP & 0.0 -- 1.4 & 0.2 & 2.76 & 0.15 & 2.0 & 1.0 $\times 10^{14}$ & 0.024 \\
		eROSITA & 0.0 -- 2.0 & 0.2 & 220 & 0.2 & $9.0$ & 3.5 $\times 10^{14}$ & 1.0 \\
		\hline
	\end{tabular}
\end{table*}

\begin{table*}[h]\centering
	\caption{\label{table:bgmnu0} Constraints on fiducial cosmology}
	\begin{tabular}{|l|c|c|c|c|c|c|c|c|c|c|c|}
		\hline
		Survey & $\sigma_\scsc{\omega_\scsc{b}}$ & $\sigma_\scsc{\omega_\scsc{c}}$ & $\sigma_\scsc{\OmegaL}$ & $\sigma_\scsc{n_\scsc{s}}$ & $\sigma_\scsc{\alpha_\scsc{s}}$ & $\sigma_\scsc{\Delta_\scsc{\mathcal{R}}^2}$ & $\sigma_\scsc{\tau}$ & $\sigma_\scsc{M\sub{\tiny lim}}$ & $\sigma_\scsc{b}$ & $\sigma_\scsc{a_1}$ & $\sigma_\scsc{a_2}$ \\
		& & & & & & $(\times 10^{-9})$ & & ($\times 10^{14} h^{-1} M_\scsc{\odot}$) & & ($h^{-1}$ Mpc) & ($h^{-2}$ Mpc$^2$) \\
		\hline\hline
		\underline{Galaxy Survey} & & & & & & & & & & & \\
		CIP & 0.0022 & 0.0086 & 0.0043 & 0.027 & 0.0082 & 0.23 & N/A & N/A & 0.0057  & 0.039 & 0.024 \\
		CIP + WMAP & 0.00026 & 0.00070 & 0.0015 & 0.0047 & 0.0037 & 0.024 & 0.0068 & N/A & 0.0055 & 0.037 & 0.021 \\
		CIP + Planck & 0.000099 & 0.00026 & 0.0012 & 0.0020 & 0.0032 & 0.013 & 0.0030 & N/A & 0.0043 & 0.029 & 0.019 \\
		SDSS + WMAP & 0.00046 & 0.0036 & 0.020 & 0.021 & 0.017 & 0.055 & 0.012 & N/A & 0.17 & 4.44 & 32.89 \\
		SDSS + Planck & 0.00013 & 0.0010 & 0.0055 & 0.0031 & 0.0049 & 0.020 & 0.0044 & N/A & 0.15 & 4.33 & 32.82 \\
		SKA & 0.00064 & 0.0028 & 0.0016 & 0.0096 & 0.0033 & 0.077 & N/A & N/A & 0.0029 & 0.028 & 0.041 \\
		SKA + WMAP & 0.00017 & 0.00061 & 0.00055 & 0.0028 & 0.0020 & 0.018 & 0.0055 & N/A & 0.0025 & 0.026 & 0.037 \\
		SKA + Planck & 0.000082 & 0.00016 & 0.00035 & 0.0016 & 0.0019 & 0.0069 & 0.0018 & N/A & 0.0022 & 0.022 & 0.032 \\
		\hline
		\underline{Cluster Survey} & & & & & & & & & & & \\
		eROSITA + WMAP & 0.00037 & 0.0021 & 0.011 & 0.013 & 0.0082 & 0.046 & 0.011 & 0.19 & N/A & N/A & N/A \\
		eROSITA + Planck & 0.00013 & 0.001 & 0.0052 & 0.003 & 0.0041 & 0.020 & 0.0043 & 0.09 & N/A & N/A & N/A \\
		SNAP + WMAP & 0.00039 & 0.0017 & 0.010 & 0.017 & 0.015 & 0.042 & 0.010 & N/A & N/A & N/A & N/A \\
		SNAP + Planck & 0.00012 & 0.00066 & 0.0036 & 0.0027 & 0.0049 & 0.018 & 0.0042 & N/A & N/A & N/A & N/A \\
		\hline
		\underline{CMB} & & & & & & & & & & & \\
		WMAP & 0.0006 & 0.0057 & 0.032 & 0.030 & 0.021 & 0.06 & 0.012 & N/A & N/A & N/A & N/A \\
		Planck & 0.00013 & 0.0011 & 0.0057 & 0.0031 & 0.0049 & 0.02 & 0.004 & N/A & N/A & N/A & N/A \\
		\hline
	\end{tabular}
\end{table*}

\begin{table}[h]\centering
	\caption{\label{table:k1mnu0} Forecasted 1-$\sigma$ marginalised uncertainties for fiducial $k\sub{\tiny IR} = 0.1 \himpc, A\sub{\tiny IR} = 0.125 \times 10^{-9}$.}
	\begin{tabular}{|l|c|c|c|c|}
		\hline
		Survey & $\sigma_\scsc{A\sub{\tiny IR}}$ & $\sigma_\scsc{k\sub{\tiny IR}}$ & $\sigma_\scsc{n_\scsc{s}}$ & $\sigma_\scsc{\alpha_\scsc{s}}$ \\
		& $(\times 10^{-9})$ & ($\himpc$) & & \\
		\hline\hline
		\underline{Galaxy Survey} & & & & \\
		CIP & 0.016 & 0.0037 & 0.027 & 0.0083 \\
		CIP + WMAP & 0.0085 & 0.0024 & 0.0047 & 0.0038 \\
		CIP + Planck & 0.0066 & 0.0015 & 0.0024 & 0.0034 \\
		SDSS + WMAP & 0.18 & 0.027 & 0.063 & 0.034 \\
		SDSS + Planck & 0.013 & 0.0023 & 0.0043 & 0.0079 \\
		SKA & 0.0039 & 0.0016 & 0.0083 & 0.0049 \\
		SKA + WMAP & 0.0032 & 0.0014 & 0.0030 & 0.0026 \\
		SKA + Planck & 0.0030 & 0.0010 & 0.0016 & 0.0023 \\
		\hline
		\underline{Cluster Survey} & & & & \\
		eROSITA + WMAP & 0.10 & 0.025 & 0.033 & 0.015 \\
		eROSITA + Planck & 0.011 & 0.0022 & 0.0040 & 0.0057 \\
		SNAP + WMAP & 0.075 & 0.027 & 0.038 & 0.025 \\
		SNAP + Planck & 0.012 & 0.0023 & 0.0038 & 0.0075 \\
		\hline
		\underline{CMB} & & & & \\
		WMAP & 0.19 & 0.027 & 0.064 & 0.035 \\
		Planck & 0.013 & 0.0023 & 0.0044 & 0.0079 \\
		\hline
	\end{tabular}
\end{table}

\begin{table}[h]\centering
	\caption{\label{table:k2mnu0} Forecasted 1-$\sigma$ marginalised uncertainties for fiducial $k\sub{\tiny IR} = 0.4 \himpc, A\sub{\tiny IR} = 0.125 \times 10^{-9}$.}
	\begin{tabular}{|l|c|c|c|c|}
		\hline
		Survey & $\sigma_\scsc{A\sub{\tiny IR}}$ & $\sigma_\scsc{k\sub{\tiny IR}}$ & $\sigma_\scsc{n_\scsc{s}}$ & $\sigma_\scsc{\alpha_\scsc{s}}$ \\
		& $(\times 10^{-9})$ & ($\himpc$) & & \\
		\hline\hline
		\underline{Galaxy Survey} & & & & \\
		CIP & 0.017 & 0.015 & 0.026 & 0.0084 \\
		CIP + WMAP & 0.0075 & 0.0070 & 0.0048 & 0.0039 \\
		CIP + Planck & 0.0072 & 0.0066 & 0.0021 & 0.0033 \\
		\hline
	\end{tabular}
\end{table}

\begin{table}[h]\centering
	\caption{\label{table:k3mnu0} Forecasted 1-$\sigma$ marginalised uncertainties for fiducial $k\sub{\tiny IR} = 1.0 \himpc, A\sub{\tiny IR} = 0.125 \times 10^{-9}$.}
	\begin{tabular}{|l|c|c|c|c|}
		\hline
		Survey & $\sigma_\scsc{A\sub{\tiny IR}}$ & $\sigma_\scsc{k\sub{\tiny IR}}$ & $\sigma_\scsc{n_\scsc{s}}$ & $\sigma_\scsc{\alpha_\scsc{s}}$ \\
		& $(\times 10^{-9})$ & ($\himpc$) & & \\
		\hline\hline
		\underline{Galaxy Survey} & & & & \\
		CIP & 0.031 & 0.096 & 0.027 & 0.0096 \\
		CIP + WMAP & 0.023 & 0.087 & 0.0046 & 0.0043 \\
		CIP + Planck & 0.022 & 0.086 & 0.0020 & 0.0036 \\
		\hline
	\end{tabular}
\end{table}
For cluster surveys, we consider an all-sky Extended ROentgen Survey with an Imaging Telescope Array\footnote{See http://www.mpe.mpg.de/heg/www/Projects/EROSITA/main.html} (eROSITA) which is a high sensitivity all-sky X-ray survey in the 0.2-12 keV energy band. The key science goal for eROSITA is to constrain the properties of dark energy using high redshift clusters of galaxies. It will have a capability to measure the spatial correlation features and evolution of a sample of about 50,000 galaxy clusters over a redshift range of 0.0 - 2.0 and will be able to find collapsed objects with mass above $3.5 \times 10^{14}\ h^{-1} M_\scsc{\odot}$. Our calculation of the expected cluster counts is in a good agreement  with the  all-sky eROSITA cluster count give in eROSITA documentation. We model the eROSITA survey by having $\Delta z = 0.2$ and $z = 0.0 - 2.0$. We also assume a 10\% prior on the mass threshold determination for the eROSITA survey. The set of parameters of eROSITA will include the mass threshold as an additional parameter reflecting the fact that we do not know accurately the mass of a cluster. The uncertainty in mass threshold will propagate into the parameter constraints. 

The SuperNova/Acceleration Probe\footnote{See http://snap.lbl.gov} (SNAP) survey \cite{Lampton2002, Aldering2005} is a space-based experiment aiming to study dark energy and dark matter. We  forecast the SNAP cluster lensing survey  constraints \cite{Marian_ea2006} (also see \cite{Fang_Haiman2007, Hamana_ea2004, Takada_Bridle2007, Wang_ea2004}). Here we take $z = 0.0 - 1.4$, $\Delta z = 0.2$, $f_{\rm sky}=0.024$, and a mass limit of $10^{14} h^{-1} M_\scsc{\odot}$. This roughly matches the number of clusters found using the more accurate selection function of \cite{Marian_ea2006} when we use their fiducial model cosmological parameters.   Also, \cite{Takada_Bridle2007} found that the signal-to-noise ratio of a more realistic selection function was about the same as taking a mass limit of $10^{14} h^{-1} M_\odot$. Our SNAP selection function is biased to slightly higher redshifts than that of \cite{Marian_ea2006}, but we expect this not to alter our predicted constraints significantly. We do not include the mass threshold into our analysis for the SNAP weak lensing survey as the weak lensing technique can be used to determined the cluster mass accurately.
 
A summary of all galaxy and cluster survey parameters is given in \tref{table:surveys}. We combined both statistical measurements from cluster number counts and the power spectrum in our predictions for the eROSITA and SNAP surveys. For the cluster surveys, we estimated the effective bias using \eq{eq:biaseff}.

We follow the work done by Barnaby \& Huang (2009) \cite{Barnaby_Huang2009}  where they found a 2-$\sigma$ constraint on the amplitude of the feature of about $2.5 \times 10^{-10}$ on scale $k \sim 0.006 - 0.1\himpc$. We conservatively define our fiducial value for the amplitude as $A\sub{\tiny IR} = 1.25 \times 10^{-10}$ and the feature at positions $k\sub{\tiny IR} = 0.1,\ 0.4\ \mbox{and}\ 1.0\ \himpc$. A survey $k\sub{max}$ limits the range of feature positions it can probe.The SDSS and SKA survey can only probe the feature at $k = 0.1 \himpc$ while CIP is the deepest survey that can probe all  of the above feature positions. The eROSITA and SNAP survey probe only the  $k = 0.1\himpc$ and $k = 0.4\himpc$ features, whereas the latter is probed by the number count component.

We define our fiducial set of parameters as the flat $\Lambda$CDM set of parameters with running of the spectral index, $\alpha_\scsc{s}$, plus the amplitude and the position of the feature $\Theta$ = ($\omega_\scsc{b}$, $\omega_\scsc{c}$, $\OmegaL$, $n_\scsc{s}$, $\alpha_\scsc{s}$, $\Delta_\scsc{\mathcal{R}}^2$, $\tau$) = (0.0227, 0.1107, 0.738, 0.969, 0.0, $2.15 \times 10^{-9}$, 0.086) and $k\sub{pivot}$ = 0.05 Mpc$^{-1}$.  Our fiducial cosmological parameters are consistent with the WMAP7 maximum likelihood values \cite{Komatsu_ea2010} although their  $k\sub{pivot}$ = 0.002 Mpc$^{-1}$. We assume a negligible neutrino mass, but we checked that adding a non-negligible mass would not significantly change our results. The constraints for the fiducial parameters without a feature are given in \tref{table:bgmnu0}. For features, we show the marginalized constraints on $A\sub{\tiny IR}$, $k\sub{\tiny IR}$, $n_\scsc{s}$ and $\alpha_\scsc{s}$  in \tref{table:k1mnu0}, \tref{table:k2mnu0}, and \tref{table:k3mnu0} for features at $k\sub{\tiny IR}$ = 0.1, 0.4 and 1.0 $\himpc$ respectively. The other cosmological parameters are included in the marginalization, but they are not particularly degenerate with the features. We use a modified version of the CAMB package\footnote{http://camb.info/} \cite{Lewis_ea2000} to generate the power spectrum for our analysis. 

\begin{figure}\centering
	\includegraphics[width=8cm]{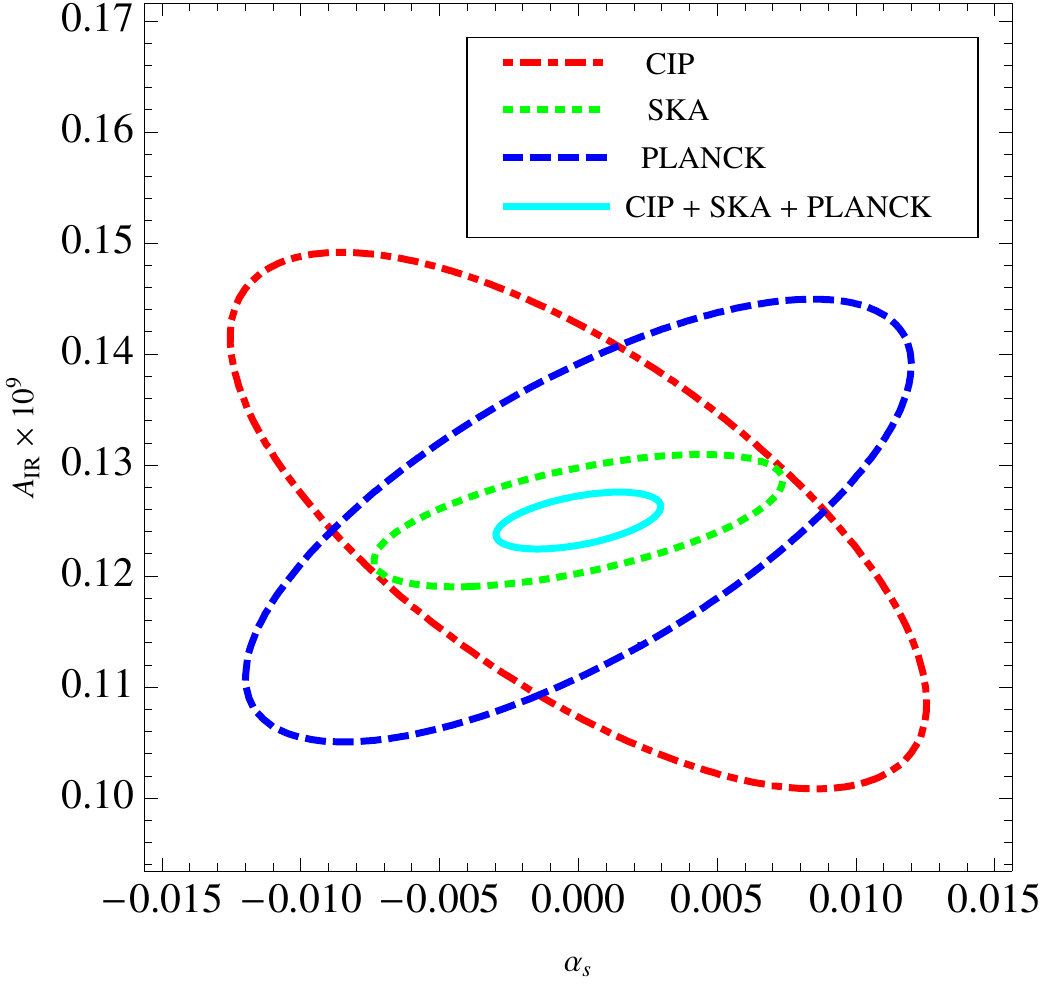}
	\caption{\label{fig:contour} Marginalised probability contours containing 68\% of the posterior probability between $\alpha_\scsc{s}$ and $A\sub{\tiny IR} = 1.25 \times 10^{-10}$.}
\end{figure}

\begin{figure}\centering
	\includegraphics[scale=0.47]{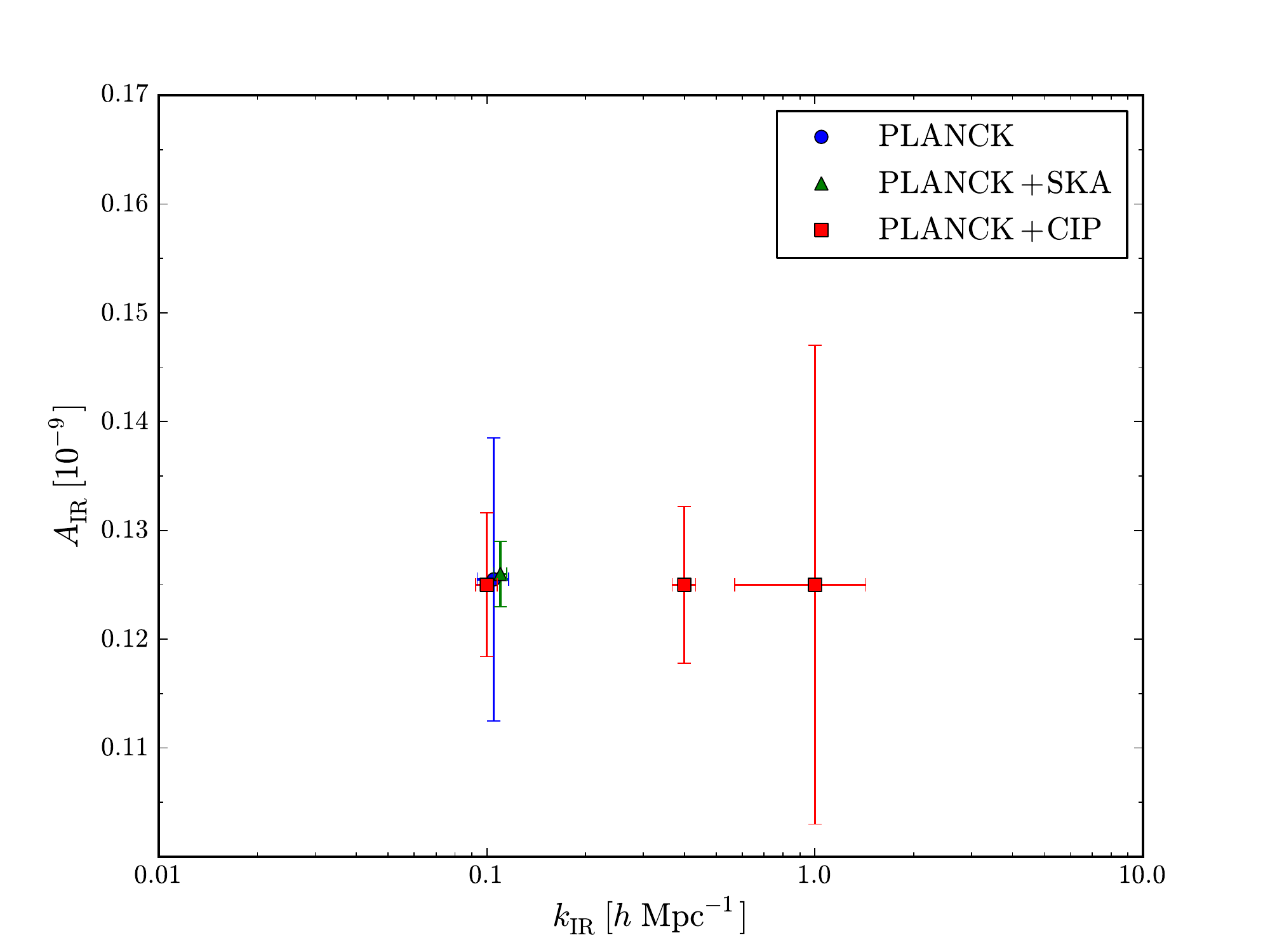}
	\caption{\label{fig:constraint} 1-$\sigma$ marginalised constraints on the amplitude and position of particle production feature for Planck, Planck + SKA and Planck + CIP. The uncertainty in $k\sub{\tiny IR}$ is multiplied by 5 to make them more visible.}
\end{figure}

\section{Discussion}
\label{sec:discussion}

%
For the standard cosmology, our cosmological constraints are in good agreement with  previous work. For example, our CIP constraints for $n_\scsc{s}$ and $\alpha_\scsc{s}$ are consistent with \cite{Colombo_ea2009, Adshead_ea2010}, our results for SDSS and SKA are consistent with \cite{Pritchard_Pierpaoli2008}. Our best constraint for ($n_\scsc{s}, \alpha_\scsc{s}$) derives from CIP + Planck and SKA + Planck which are (0.0020, 0.0032) and (0.0016, 0.0019) respectively. However, since CIP and SKA surveys probe the power spectrum at exclusively different redshift range, we can consider them as two independent surveys. The combined CIP + SKA + Planck improves the constraints to (0.0014, 0.0017).


From \tref{table:k1mnu0} -- \ref{table:k3mnu0}, we investigate the effect of a feature on cosmological parameters on scales $k = 0.1, 0.4$ and 1.0 $\himpc$ respectively. The WMAP+SDSS constraints for the $0.1\himpc$ feature are consistent with those found for the actual data \cite{Barnaby_Huang2009}. As the CMB is limited to $\ell \leq 2000$, it can only directly constrain the  $k = 0.1\himpc$ feature. It may be possible to extend this range somewhat for polarization, but probably not enough to completely encompass the width of the 0.4 $\himpc$ centred feature. The cluster surveys were only able to help directly constrain the feature at 0.1 $\himpc$. Although in previous work \cite{Chantavat_ea2009}, we found that cluster surveys could constrain the primordial power spectrum directly at scales of 0.4$\himpc$, this was for a linear piecewise binning of the primordial power spectrum where  the location of the feature did not need to be constrained.

With features at different positions, the constraint on $n_\scsc{s}$ remains almost the same while the constraints on $\alpha_\scsc{s}$ are worsened by about a factor of 2. The degeneracies between $A\sub{\tiny IR}$ and $k\sub{\tiny IR}$ have a significant impact on the running as a variation in $\alpha_\scsc{s}$ could be made on a more localised scale than a variation in $n_\scsc{s}$. However, there is more of an improvement on $\alpha_\scsc{s}$ with CIP + Planck. \fref{fig:contour} shows that CIP could be used to break the degeneracy between $A\sub{\tiny IR}$ and $\alpha_\scsc{s}$ while SKA only provides a tightened constraint but the degeneracy is still in the same direction. CIP is in a different direction as it has most of its constraining power at higher $k$ values and so the degeneracy between $A\sub{\tiny IR}$, $\Delta_\scsc{\cal R}^{2}$, and $\alpha_\scsc{s}$ is different.

Our inference of $\sigma_\scsc{A\sub{\tiny IR}}$ and $\sigma_\scsc{k\sub{\tiny IR}}$ is summarised in \fref{fig:constraint}. As can be seen, Planck will improve the constraint at the $0.1\himpc$ scale by about a factor of about 10 in comparison to WMAP. The addition of SKA further improves the constrain by another factor of about 5. On wavenumbers up to about $1.0 \himpc$ CIP combined with Planck could provide similar constraints.

Even smaller scales are constrainable, but then CIP would not be able to probe the full extent of the feature. Also, it may be possible to further improve the constraints by including the non-Gaussianity associated with the feature \cite{Barnaby2010}.

\begin{figure}\centering
	\includegraphics[scale=0.45]{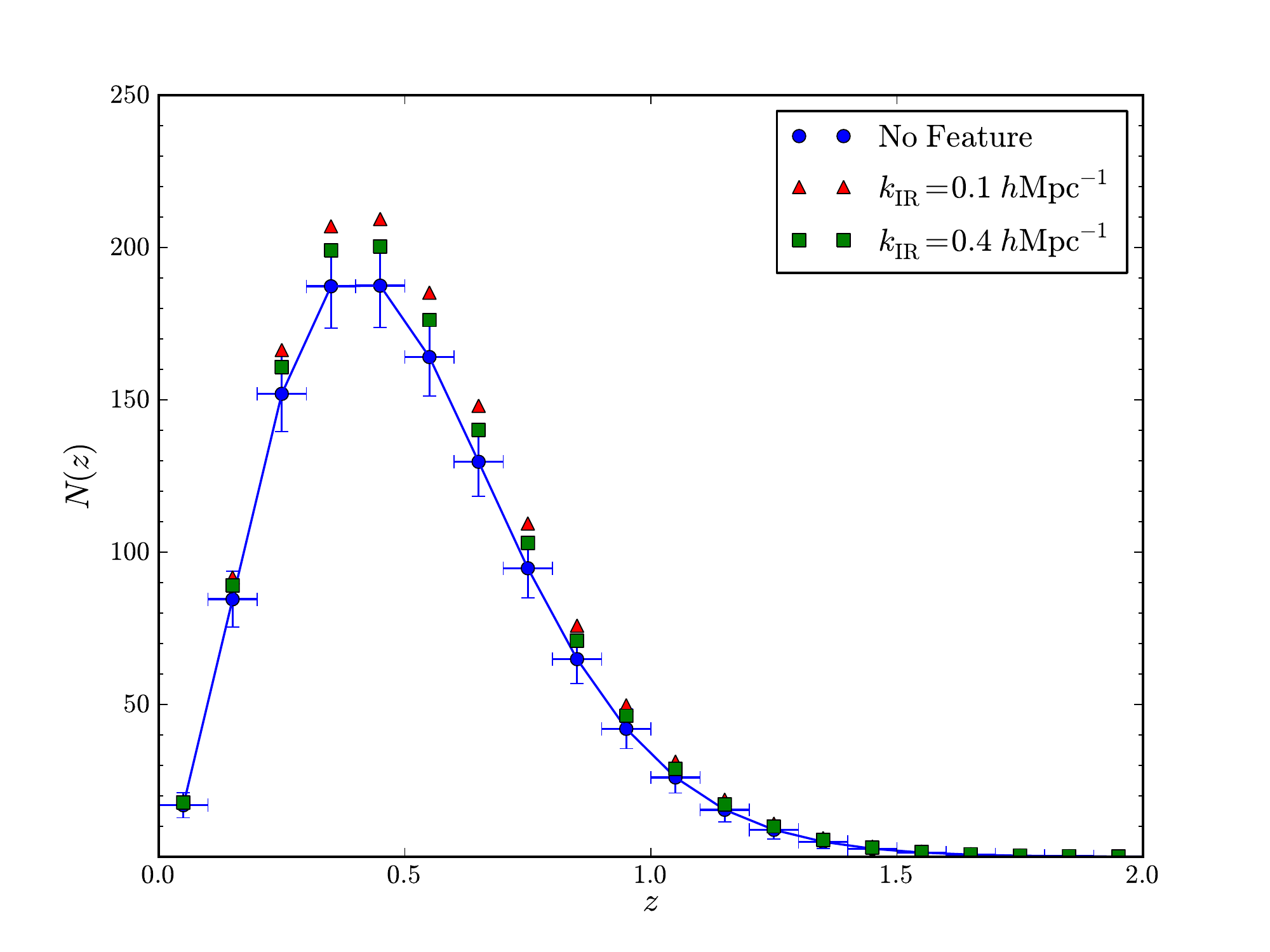}
	\caption{\label{fig:numbercounts} The effect of a change in primordial power spectrum on the eROSITA all-sky survey. The circle dots represent number of cluster that would be observed by eROSITA for no feature power spectrum. The triangle and square dots show excess in number count for feature at $k\sub{\scriptsize IR} = 0.1$ and 0.4 $\himpc$ respectively.
	The error bars are $1\sigma$.}
\end{figure}

\begin{figure}\centering
	\includegraphics[scale=0.5]{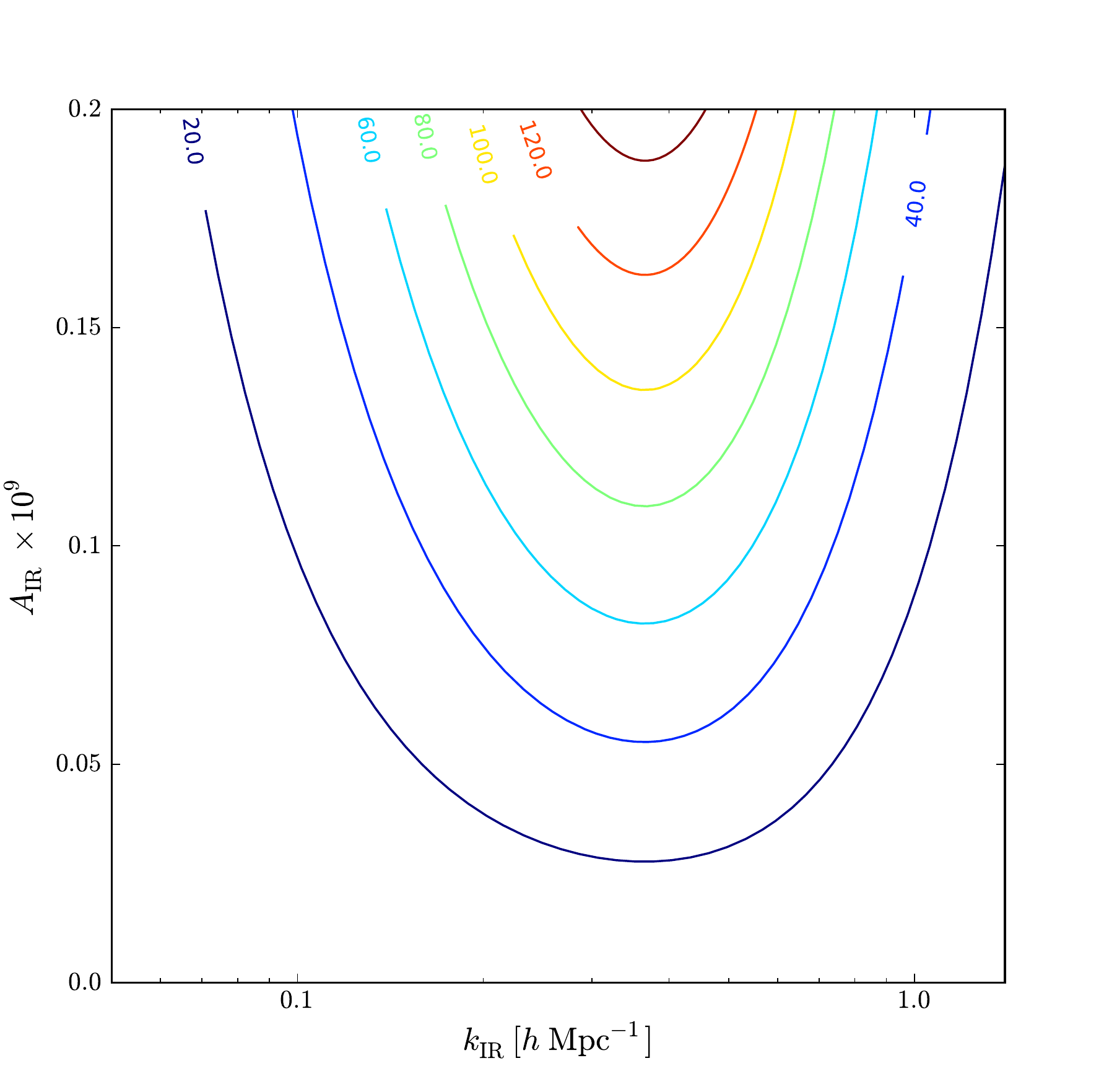}
	\caption{\label{fig:AmpKposContour} The excess number count from SNAP survey due to the feature with an ampitude $A\sub{\tiny IR}$ and position at $k\sub{\tiny IR}$.}
\end{figure}

{  Even though the excess in cluster number count can be easily obtained from cluster surveys (See \fref{fig:numbercounts} for eROSITA), they do not give better constraints in comparison to galaxies surveys partly due to low number statistics}. In addition, the cluster number counts are not good at simultaneous determining the amplitude of the features and the positions due to the degeneracy in cluster counts. Because the number of clusters is determined by the integral of $P(k)$ (See \eq{eq:massfunction} and \eq{eq:sigma}) the features with different amplitude and position can conspire to yield the same integral, hence, number counts. \fref{fig:AmpKposContour} shows the contour plot of excesses in number counts for different amplitude and position of the features.

In summary, we have demonstrated that future surveys can potentially probe the primordial power spectrum, for particle production-induced features, significantly more accurately and to significantly smaller length scales than at present. Even if no features are detected, at least the simplest models will have been tested much more precisely and on a much greater range of scales.

\section{Acknowledgement}

TC is funded by the Institute for the Promotion of Teaching Science and Technology (IPST) in Thailand. CG is funded by the Beecroft Institute for
Particle Astrophysics and Cosmology.


\bibliography{IRbib}

\label{lastpage}

\end{document}